Quantum Networking and Communications

IEEE Transactions on Quantum Engineering

Received January 16, 2020; revised February 26, 2020; accepted February 28, 2020; date of publication April 22, 2020; date of current version May 18, 2020.

Digital Object Identifier 10.1109/TQE.2020.2978454
# Improved Gilbert–Varshamov Bound for Entanglement-Assisted Asymmetric Quantum Error Correction by Symplectic Orthogonality

RYUTAROH MATSUMOTO[1,2]  (Member, IEEE)

[1]Department of Information and Communication Engineering, Nagoya University, Nagoya 464-8603, Japan
[2]Department of Mathematical Sciences, Aalborg University, Aalborg 9100, Denmark
This work was supported by the Japan Society for the Promotion of Science under Grant 17K06419.
**ABSTRACT** We propose and prove an existential theorem for entanglement-assisted asymmetric quantum error correction. Then, we demonstrate its superiority over the conventional one.

## I. INTRODUCTION

Quantum error-correcting codes (QECC) are important for construction of quantum computers, as the fault-tolerant quantum computation is based on QECC [13]. There are two kinds of typical errors in quantum information, one is called a bit error and the other is called a phase error. Steane [15] first studied the asymmetry between probabilities of the bit and the phase errors, and he also considered QECC for asymmetric quantum errors, which are called asymmetric QECC. Research on asymmetric QECC has become very active recently; see [6], [8], [15], and the references therein.

Most of QECCs, for both asymmetric and symmetric error correction, come from classical codes. Following [11], one can obtain QECCs of length $n$ over a finite field $\mathbf{F}_q$ of order $q$ from additive codes included in $\mathbf{F}_q^{2n}$, which are self-orthogonal with respect to a trace symplectic form. Working on this construction, QECCs of length $n$ over $\mathbf{F}_q$ can be derived from classical self-orthogonal codes with respect to the Hermitian inner product included in $\mathbf{F}_{q^2}^n$, and also from codes in $\mathbf{F}_q^n$ that are self-orthogonal with respect to the Euclidean inner product.

The previously mentioned self-orthogonality conditions (or some similar requirements of inclusion of codes in the dual of others) prevent the usage of many common classical codes for providing quantum codes. Brun *et al.* in [2] proposed to share entanglement between encoder and decoder to simplify the theory of quantum error correction and increase the communication capacity. With this new formalism, entanglement-assisted quantum stabilizer codes can be constructed from any classical linear code giving rise to entanglement-assisted quantum error-correcting codes (EAQECCs). Recently, nonbinary generalization of EAQECCs was studied [4], [10]. On the other hand, until very recently, no researcher had studied entanglement-assisted asymmetric quantum error correction, though its necessity and importance seem pretty obvious at hindsight.

Galindo *et al.* [5] recently introduced asymmetric error correction to EAQECCs, and also proved a Gilbert–Varshamov (GV) type existential theorem of codes with a given set of parameters. They used the direct products of two linear spaces and the Euclidean inner product. The GV-type existential theorems reveal the optimal performances of both classical and quantum error correction, and many variations of the GV-type existential theorems have been studied [14]. In particular, a GV-type existential theorem was proved for binary EAQECC for symmetric errors in [9].

In this article, we propose a different version of the GV bound that uses the symplectic inner product and linear spaces, not necessarily direct products. Since our proposed GV bound covers a wider class of linear spaces, it gives a better existential bound than [5], as demonstrated by the examples in Table 1. Therefore, this article sheds new light on the optimal entanglement-assisted asymmetric quantum error correction. This article is organized as follows. In Section II, we propose a new existential theorem for entanglement-assisted asymmetric quantum error correction. In Section III,

VOLUME 1, 2020
This work is licensed under a Creative Commons Attribution 4.0 License. For more information, see http://creativecommons.org/licenses/by/4.0/
4100604

**TABLE 1.** Comparison between the proposed theorem 2 and the conventional theorem 4 for the parameters in [5, Table I]

| $q$ | $n$ | $k_1$ | $k_2$ | $c$ | $P_{\text{old}}$ [5] | $P_{\text{new}}$ (Proposed) |
|---|---|---|---|---|---|---|
| 4 | 15 | 3 | 1 | 1 | (2, 1) | (2, 1), (1, 2) |
| 5 | 24 | 5 | 3 | 3 | (2, 2) | (4, 1), (2, 2), (1, 4) |
| 7 | 19 | 7 | 4 | 4 | (4, 2) | (7, 1), (5, 2), (4, 3), (3, 4), (2, 5), (1, 7) |
| 7 | 19 | 13 | 10 | 10 | (8, 6) | (19, 2), (16, 3), (12, 4), (10, 5), (9, 6), (8, 7), (7, 8), (6, 9), (5, 10), (4, 12), (3, 16), (2, 19) |
| 8 | 63 | 7 | 1 | 1 | (3, 1) | (4, 1), (2, 2), (1, 4) |
| 8 | 63 | 11 | 3 | 3 | (5, 2) | (6, 1), (5, 2), (4, 3), (3, 4), (2, 5), (1, 6) |
| 9 | 40 | 10 | 5 | 5 | (5, 3) | (8, 1), (6, 2), (5, 3), (4, 4), (3, 5), (2, 6), (1, 8) |
| 9 | 40 | 12 | 3 | 3 | (6, 2) | (8, 1), (6, 2), (5, 3), (4, 4), (3, 5), (2, 6), (1, 8) |
| 9 | 40 | 12 | 7 | 7 | (6, 3) | (11, 1), (9, 2), (7, 3), (6, 4), (5, 5), (4, 6), (3, 7), (2, 9), (1, 11) |
| 16 | 51 | 9 | 3 | 3 | (5, 2) | (7, 1), (5, 2), (4, 3), (3, 4), (2, 5), (1, 7) |
| 16 | 51 | 11 | 1 | 1 | (6, 1) | (7, 1), (5, 2), (4, 3), (3, 4), (2, 5), (1, 7) |
| 16 | 51 | 11 | 3 | 3 | (6, 2) | (8, 1), (6, 2), (5, 3), (4, 4), (3, 5), (2, 6), (1, 8) |
| 16 | 51 | 17 | 5 | 5 | (10, 3) | (13, 1), (11, 2), (10, 3), (9, 4), (8, 5), (6, 6), (5, 8), (4, 9), (3, 10), (2, 11), (1, 13) |
| 16 | 51 | 19 | 5 | 5 | (11, 3) | (15, 1), (13, 2), (11, 3), (10, 4), (9, 5), (8, 6), (7, 7), (6, 8), (5, 9), (4, 10), (3, 11), (2, 13), (1, 15) |
| 16 | 51 | 23 | 3 | 3 | (14, 2) | (16, 1), (14, 2), (13, 3), (11, 4), (10, 5), (9, 6), (8, 7), (7, 8), (6, 9), (5, 10), (4, 11), (3, 13), (2, 14), (1, 16) |
| 16 | 51 | 23 | 9 | 9 | (14, 5) | (21, 1), (19, 2), (17, 3), (16, 4), (14, 5), (13, 6), (12, 7), (11, 8), (10, 9), (9, 10), (8, 11), (7, 12), (6, 13), (5, 14), (4, 16), (3, 17), (2, 19), (1, 21) |
| 16 | 51 | 27 | 5 | 5 | (17, 3) | (21, 1), (19, 2), (17, 3), (16, 4), (14, 5), (13, 6), (12, 7), (11, 8), (10, 9), (9, 10), (8, 11), (7, 12), (6, 13), (5, 14), (4, 16), (3, 17), (2, 19), (1, 21) |
| 25 | 48 | 6 | 4 | 4 | (4, 2) | (6, 1), (5, 2), (3, 3), (2, 5), (1, 6) |
| 25 | 48 | 10 | 4 | 4 | (6, 2) | (8, 1), (7, 2), (6, 3), (5, 4), (4, 5), (3, 6), (2, 7), (1, 8) |
| 25 | 48 | 10 | 7 | 7 | (6, 4) | (11, 1), (9, 2), (8, 3), (7, 4), (5, 5), (4, 7), (3, 8), (2, 9), (1, 11) |
| 25 | 48 | 12 | 3 | 3 | (7, 2) | (9, 1), (8, 2), (6, 3), (5, 4), (4, 5), (3, 6), (2, 8), (1, 9) |
| 25 | 48 | 12 | 6 | 6 | (7, 4) | (11, 1), (10, 2), (8, 3), (7, 4), (6, 5), (5, 6), (4, 7), (3, 8), (2, 10), (1, 11) |

we compare our proposal with the conventional one [5]. Finally, concluding remarks are given in Section IV.

## II. PROPOSED GV BOUNDS

Galindo *et al.* [4], [5] provided the following construction of entanglement-assisted asymmetric quantum error correction in their two subsequent papers.

*Theorem 1 (see [4], [5]):* Let $C \subseteq \mathbf{F}_q^{2n}$ be an $\ell$-dimensional $\mathbf{F}_q$-linear space and $H = (H_X|H_Z)$ an $\ell \times 2n$ matrix whose row space is $C$. Denote by $c$ the minimum required number of maximally entangled quantum states in $\mathbf{C}^q \otimes \mathbf{C}^q$. Then,

$$2c = \text{rank}\left(H_X H_Z^T - H_Z H_X^T\right) = \dim C - \dim\left(C \cap C^{\perp_s}\right),$$

and the QECC defined by $C$ encodes $n - \ell + c$ qudits in $\mathbf{C}^q$ into $n$ qudits. The constructed EAQECC can detect $d_x$ bit errors and $d_z$ phase errors, if $w_H(\vec{v}_x) \leq d_x - 1$ and $w_H(\vec{v}_z) \leq d_z - 1$ implies $(\vec{v}_x|\vec{v}_z) \notin (C^{\perp_s} \setminus (C \cap C^{\perp_s}))$ for any $\vec{v}_x \in \mathbf{F}_q^n$ and $\vec{v}_z \in \mathbf{F}_q^n$, where $w_H(\cdot)$ denotes the Hamming weight of a vector [14]. In sum, $C$ provides an $[[n, n - \ell + c, d_z/d_x; c]]_q$ EAQECC over the field $\mathbf{F}_q$.

Galindo *et al.* [5] provided a GV-type bound only when the above $C$ can be written as $C_1 \times C_2$, where $C_1, C_2 \subset \mathbf{F}_q^n$. We remove the limitation $C = C_1 \times C_2$ and propose the following new theorem. The comparison between the proposed and the conventional ones will be given in the following section.

*Theorem 2:* Assume the existence of integers $n \geq 1$, $1 \leq \ell < n + c$, $d_x \geq 1$, $d_z \geq 1$, $0 \leq c \leq \ell/2$ such that

$$\frac{q^{2n-\ell} - q^{\ell-2c}}{q^{2n} - 1}\left(\sum_{i=0}^{d_x-1}\binom{n}{i}(q-1)^i \sum_{j=0}^{d_z-1}\binom{n}{i}(q-1)^j - 1\right)$$

$$< 1. \quad (1)$$

Then, there exists an $\mathbf{F}_q$-linear code $C \subseteq \mathbf{F}_q^{2n}$ such that $\dim C = \ell$, $\dim C - \dim(C^{\perp_s} \cap C) = 2c$, and for any $\vec{v}_x$ and $\vec{v}_z$ with $w_H(\vec{v}_x) \leq d_x - 1$ and $w_H(\vec{v}_z) \leq d_z - 1$ we have $(\vec{v}_x, \vec{v}_z) \notin (C^{\perp_s} \setminus (C \cap C^{\perp_s}))$. This means the existence of an $[[n, n - \ell + c, d_z/d_x; c]]_q$ EAQECC over the field $\mathbf{F}_q$.

*Proof:* We will use an argument similar to the proof of the GV bound for stabilizer codes [3]. Let $\text{Sp}(q, n)$ be the symplectic group over $\mathbf{F}_q^{2n}$ [7, Sec. 3] and $A(\ell, c)$ the set of $\mathbf{F}_q$-linear spaces $V \subseteq \mathbf{F}_q^{2n}$ such that $\dim V = \ell$ and

$$\dim V - \dim\left(V^{\perp_s} \cap V\right) = 2c.$$

For $\vec{0} \neq \vec{e} \in \mathbf{F}_q^{2n}$, define

$$B(\ell, c, \vec{e}) = \left\{V \in A(\ell, c) \mid \vec{e} \in V^{\perp_s} \setminus (V^{\perp_s} \cap V)\right\}.$$

Taking into account that the symplectic group acts transitively on $\mathbf{F}_q^{2n} \setminus \{\vec{0}\}$ [1], [7], it holds that for nonzero $\vec{e}_1, \vec{e}_2 \in \mathbf{F}_q^{2n}$, there exists $M \in \text{Sp}(q, n)$ such that $\vec{e}_1 M = \vec{e}_2$, and, for $V_1, V_2 \in A(\ell, c)$, there exists $M \in \text{Sp}(q, n)$ such that $V_1 M = V_2$.

Therefore, for nonzero elements $\vec{e}_1, \vec{e}_2 \in \mathbf{F}_q^{2n}$ with $\vec{e}_1 M_1 = \vec{e}_2$ ($M_1 \in \text{Sp}(q, n)$) and some fixed linear space $V_1 \in A(\ell, c)$, we have the following chain of equalities:

$\text{card}(B(\ell, c, \vec{e}_1))$

$= \text{card}(\{V \in A(\ell, c) \mid \vec{e}_1 \in V^{\perp_s} \setminus (V^{\perp_s} \cap V)\})$

$= \text{card}(\{V_1 M \mid \vec{e}_1 \in V_1^{\perp_s} M \setminus (V_1^{\perp_s} M \cap V_1 M), M \in \text{Sp}(q, n)\})$

$= \text{card}(\{V_1 M M_1^{-1} \mid \vec{e}_1 \in V_1^{\perp_s} M M_1^{-1} \setminus$
$(V_1^{\perp_s} M M_1^{-1} \cap V_1 M M_1^{-1}), M \in \text{Sp}(q, n)\})$

$= \text{card}(\{V_1 M \mid \vec{e}_1 M_1 \in V_1^{\perp_s} M \setminus (V_1^{\perp_s} M \cap V_1 M),$
$M \in \text{Sp}(q, n)\})$

$$= \text{card}(\{V_1 M \mid \vec{e}_2 \in V_1^{\perp_s} M \setminus (V_1^{\perp_s} M \cap V_1 M),$$
$$M \in \text{Sp}(q, n)\})$$
$$= \text{card}(\{V \in A(\ell, c) \mid \vec{e}_2 \in V^{\perp_s} \setminus (V^{\perp_s} \cap V)\})$$
$$= \text{card}(B(\ell, c, \vec{e}_2))$$

where card denotes the cardinality of a set.

For each $V \in A(\ell, c)$, the number of vectors $\vec{e}$ in $\mathbf{F}_q^{2n}$ such that $\vec{e} \in V^{\perp_s} \setminus (V^{\perp_s} \cap V)$ is

$$\text{card}(V^{\perp_s}) - \text{card}(V^{\perp_s} \cap V) = q^{2n-\ell} - q^{\ell-2c}.$$

The number of pairs $(\vec{e}, V)$ such that $\vec{0} \neq \vec{e} \in V^{\perp_s} \setminus (V^{\perp_s} \cap V)$ is

$$\sum_{\vec{0} \neq \vec{e} \in \mathbf{F}_q^{2n}} \text{card}(B(\ell, c, \vec{e})) = \text{card}(A(\ell, c))\left(q^{2n-\ell} - q^{\ell-2c}\right)$$

which implies

$$\frac{\text{card}(B(\ell, c, \vec{e}))}{\text{card}(A(\ell, c))} = \frac{q^{2n-\ell} - q^{\ell-2c}}{q^{2n} - 1}. \quad (2)$$

If there exists $V \in A(\ell, c)$ such that $V \notin B(\ell, c, \vec{e})$ for all nonzero vectors $\vec{e} = (\vec{e}_x | \vec{e}_z)$ such that $0 \leq w_H(\vec{e}_x) \leq d_x - 1$ and $0 \leq w_H(\vec{e}_z) \leq d_z - 1$, then there will exist $V$ with the desired properties. The number of nonzero vectors $\vec{e} = (\vec{e}_x | \vec{e}_z)$ such that $0 \leq w_H(\vec{e}_x) \leq d_x - 1$ and $0 \leq w_H(\vec{e}_z) \leq d_z - 1$ is given by

$$\sum_{i=0}^{d_x-1} \binom{n}{i}(q-1)^i \times \sum_{j=0}^{d_z-1} \binom{n}{j}(q-1)^j - 1. \quad (3)$$

By combining (2) and (3), we see that inequality (1) is a sufficient condition for ensuring the existence of a code $C$ as in our statement. ∎

In the coding theory, asymptotic optimality has been investigated [14], as long code length is sometimes required for higher error correction capability. By following an argument similar to [12], one can easily deduce an asymptotic version of Theorem 2 as follows.

*Corollary 3:* Let $L$, $\delta_x$, $\delta_z$, and $\lambda$ be real numbers such that $0 \leq L \leq 1 + \lambda$, $0 \leq \delta_x < 1/2$, $0 \leq \delta_z < 1/2$, and $0 \leq \lambda \leq L/2$. Let $h(x) := -x \log_q x - (1-x) \log_q(1-x)$ be the $q$-ary entropy function. For $n$ sufficiently large, the inequality

$$h(\delta_x) + \delta_x \log_q(q-1) + h(\delta_z) + \delta_z \log_q(q-1) < L \quad (4)$$

implies the existence of a code $C \subseteq \mathbf{F}_q^{2n}$ over $\mathbf{F}_q$ such that $\dim C = \lceil nL \rceil$, $\dim C - \dim(C^{\perp_s} \cap C) = \lceil 2n\lambda \rceil$, and for any $\vec{v}_x$ and $\vec{v}_z$ with $w_H(\vec{v}_x) \leq n\delta_x - 1$ and $w_H(\vec{v}_z) \leq n\delta_z - 1$ we have $(\vec{v}_x, \vec{v}_z) \notin (C^{\perp_s} \setminus (C \cap C^{\perp_s}))$. This means that there exists $[[n, n(1-L+\lambda), n\delta_z/n\delta_x; n\lambda]]_q$ EAQECC.

## III. COMPARISON BETWEEN THE PROPOSED AND THE CONVENTIONAL BOUNDS

In this section, we compare our proposed bound Theorem 2 with the conventional one.

*Theorem 4 (see [5]):* Consider the positive integer numbers $n$, $k_1$, $k_2$, $d_z$, $d_x$, and $c$ such that $k_1 \leq n$, $k_2 \leq n$ and

$$k_1 + k_2 - n \leq c \leq \min\{k_1, k_2\}$$

which satisfy the following inequality:

$$\frac{q^{n-k_1} - q^{k_2-c}}{q^n - 1} \sum_{i=1}^{d_z-1} \binom{n}{i}(q-1)^i$$
$$+ \frac{q^{n-k_2} - q^{k_1-c}}{q^n - 1} \sum_{i=1}^{d_x-1} \binom{n}{i}(q-1)^i < 1 \quad (5)$$

then there exists an $[[n, n - k_1 - k_2 + c, d_z/d_x; c]]_q$ asymmetric EAQECC.

The parameter $\ell$ in Theorem 2 corresponds to $k_1 + k_2$ in Theorem 4.

In order to compare Theorems 2 and 4, we choose the same $q$, $n$, $c$, and $\ell = k_1 + k_2$ from each line in [5, Table 1]. For a tuple of $q$, $n$, $c$ and $\ell = k_1 + k_2$, they [5, Sec. VI] also introduced the set $P$ in order to quantify the maximum possible distances for a given parameters $q$, $n$, $c$, and $\ell = k_1 + k_2$ with which existence of a quantum code is ensured by a version of GV bounds. Specifically, for fixed values $(q, n, k_1, k_2, c)$ (or $(q, n, \ell = k_1 + k_2, c)$), we consider the set $P_\text{old}$ (or $P_\text{new}$) of pairs $(d_1, d_2)$ of Z-minimum and X-minimum distances of asymmetric EAQECCs such that $(d_1, d_2)$ satisfies inequality (5) [or inequality (1)] but either $(d_1 + 1, d_2)$ or $(d_1, d_2 + 1)$ violates the inequality (5) [or inequality (1)], respectively. If for any $(d_1, d_2) \in P_\text{old}$ there exists $(d_1', d_2') \in P_\text{new}$ such that $d_1 \leq d_1'$ and $d_2 \leq d_2'$, then we can say that the proposed Theorem 2 *improves* Theorem 4. In Table 1, we see that it is actually the case.

## IV. CONCLUSION

In this article, we studied what is the optimal entanglement-assisted asymmetric quantum error correction. Specifically, we proposed a GV-type existential theorem. Then, we compared our proposal with the conventional bound in [5] and demonstrated its superiority over [5] in Table 1. The difference between paper [5] and our proposal is that we impose no restriction on linear spaces used for code construction, whereas paper [5] assumed that the linear spaces were direct products. We did not consider explicit code constructions when the linear spaces were not restricted to the direct products, which is a future research agenda.

## ACKNOWLEDGMENT

The author would like to thank anonymous reviewers for their careful reading valuable comments pointing out confusing errors in the initial manuscript.

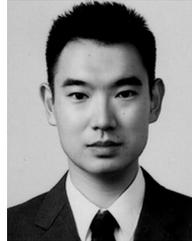

**RYUTAROH MATSUMOTO** (Member, IEEE) was born in Nagoya, Japan, on November 29, 1973. He received the B.E. degree in computer science, the M.E. degree in information processing, and the Ph.D. degree in electrical and electronic engineering from Tokyo Institute of Technology, Tokyo, Japan, in 1996, 1998, and 2001, respectively.

He was an Assistant Professor from 2001 to 2004, and was an Associate Professor from 2004 to 2017 with the Department of Information and Communications Engineering, Tokyo Institute of Technology. Since 2017, he has been an Associate Professor with the Department of Information and Communication Engineering, Nagoya University, Nagoya, Japan. In 2011 and 2014, he was as a Velux Visiting Professor with the Department of Mathematical Sciences, Aalborg University, Aalborg, Denmark. His research interests include error-correcting codes, quantum information theory, information theoretic security, and communication theory.

Dr. Matsumoto was the recipient of the Young Engineer Award from IEICE and the Ericsson Young Scientist Award from Ericsson Japan in 2001. He was also the recipient of the Best Paper Awards from IEICE in 2001, 2008, 2011, and 2014.